\documentclass{PoS}

\title{New Results on Solar Neutrinos}

\ShortTitle{New Results on Solar Neutrinos}

\author{\speaker{A. Bellerive}\thanks{On behalf of the SNO Collaboration.}\\
        Ottawa-Carleton Institute for Physics \\
Department of Physics \\
Carleton University \\
1125 Colonel By Drive,Ottawa, K1S 5B6, Canada\\
        E-mail: \email{alain\_bellerive@carleton.ca}}

\abstract{This paper 
reviews the constraints on the solar
neutrino mixing parameters with data collected by the Homestake, 
SAGE, GALLEX, Kamiokande, SuperKamiokande, Borexino and SNO experiments.
An emphasis will be given to the global solar neutrino analyses
in terms of matter-enhanced oscillation of two and three 
active flavors.
The results to-date, including both solar model dependent and 
independent measurements, indicate that electron neutrinos
are changing to other active types on route to the Earth 
from the Sun.
The total flux of solar neutrinos is found
to be in very good agreement with solar model calculations. 
Today, solar neutrino measurements focus on greater accuracy for mixing parameters
and on better sensitivity to low neutrino energies. This article also summarizes
near future prospects in the field of solar neutrino physics.
}

\FullConference{35th International Conference of High Energy Physics - ICHEP2010,\\
		July 22-28, 2010\\
		Paris France}

\begin{document}

\section{Introduction}

The deficit of neutrinos detected coming from the Sun compared with our 
expectations based on laboratory measurements, 
known as the Solar Neutrino Problem, has remained one of the outstanding 
problems in basic physics for over thirty years. 
It appeared inescapable that either our understanding of the energy producing 
processes in the Sun is seriously defective, 
or neutrinos, some of the fundamental particles in the Standard Model, have 
important properties which have yet to be identified. 
It was indeed argued by some that we needed to change our ideas on how energy 
was produced in fusion reactions inside the Sun. 
Others suggested that the problem arose due to peculiar characteristics of 
neutrinos such as oscillations and matter effects. 
It is then useful to review the evolution of our understanding from the data 
collected by various solar neutrino experiments. 
For completeness, near term prospects are included in the 
discussion presented here.

\section{Solar Neutrinos}

The detailed prediction of the electron neutrino flux created by the thermonuclear reactions 
in the interior of 
the Sun was performed by John Bahcall and his collaborators (for example see~\cite{bib:BP}). 
Their calculations are referred to as the Standard Solar Model (SSM). 
In this paper, the SSM calculations are used
to compare experimental results and theoretical predictions. 
The relevant point for the discussion that follows is that SSM theoretical 
uncertainties on many solar fluxes have been reduced~\cite{bib:BPS08}.
As a result, the errors on the $^7Be$, $pep$, $^8B$ and $hep$ fluxes are now 
$\pm 6\%$, $\pm 1.1\%$, $\pm 11\%$ and $\pm 16\%$, respectively.

\section{Chlorine Experiment}

The exploration of solar neutrinos started in the mid-1960's with 
Ray Davis~\cite{bib:cl}. It led
to the first experiment that successfully detected neutrinos coming from the Sun. The 
experiment of Davis
and his team was carried out deep underground in the Homestake mine in the US. 
The detector was based on a concept first proposed by Bruno Pontecorvo at Chalk 
River in 1946, in which 
neutrino reactions on chlorine are measured. Neutrinos striking chlorine can make an 
isotope of argon through the 
reaction
\[
\nu_e \,+\, ^{37}Cl \to e^- \,+\, ^{37}Ar \, ,
\]
with an energy threshold of 0.814~MeV. The first results 
were announced in 1968. 

The chlorine experiment took data until 1995 and clearly showed argon atoms 
produced by neutrinos, but the predicted production rate  was four times the measured
value~\cite{bib:cleveland}:
\begin{equation}
\Phi_{\rm{Cl}} = 2.56 \pm 0.23 \mbox{~SNU} \, ,
\end{equation}
while the SSM predicted rate was about $\Phi_{\rm{Cl}}({\rm{SSM}}) = 7.6$~SNU.
A SNU (Solar Neutrino Unit) is the product of the solar neutrino 
fluxes (measured or calculated) and the calculated cross sections. Hence
one SNU equals one capture per second and per $10^{36}$ target atoms.

\section{Gallium Experiments}

While the chlorine detector was mainly sensitive to the highest energy 
neutrinos, 
two gallium experiments, one at the Baksan laboratory~\cite{bib:sage} in Russia and 
one at the Gran Sasso laboratory~\cite{bib:gno} in Italy,
were set up to test the oscillation hypothesis at lower energy. 
Like the $^{37}Cl$ detector, the gallium detectors 
could only detect electron type neutrinos because they looked for the reaction
\[
\nu_e \,+\, ^{71}Ga \to e^- \,+\, ^{71}Ge \, .
\]
The energy threshold of the $^{71}Ga$ detectors is 0.233~MeV and hence allows the 
interaction of
$pp$, $^7Be$, $^8B$, and $pep$ neutrinos. The Russian-American 
group (SAGE) used a liquid metal target which 
contained 50 tons of gallium; while the European group (GALLEX/GNO) used 30 tons 
of natural gallium in an   
aqueous acid solution. Small proportional counters are used to count the germanium 
from the radiochemical target.
The $^{71}Ge$ electron capture decay occurs with a mean-life of 16.5 days.
The Auger electrons and X-rays produce the typical L-peak and K-peak energy
distribution. As a cross-check, both peaks
are counted separately. Both experiments found about half of the expected rate. The most recent 
results of SAGE and GALLEX/GNO yield~\cite{bib:ga}
\begin{equation}
\Phi_{\rm{Ga}} = 66.1 \pm 3.1 \mbox{~SNU} \, .
\end{equation}
The data are incompatible with the SSM since 
the expected rate is about $\Phi_{\rm{Ga}}({\rm{SSM}}) = 129$~SNU.

\section{Kamiokande and SuperKamiokande}

Following the first observations from the Cl experiment the 
first priority
was obviously an experimental confirmation of the solar-neutrino 
deficit. This was 
provided in 1987 by the Kamiokande water \v{C}erenkov 
detector~\cite{bib:kamioka} in Japan, which
also saw a significant (but, interestingly enough, not an identical) suppression 
of the measured rate of neutrinos from the Sun.
The Kamiokande Collaboration demonstrated that the neutrinos are
actually coming from the direction of the Sun by reconstructing the direction 
of flight of the incident neutrinos from the neutrino-electron 
scattering (ES) reaction $\nu_x \,+\, e^- \to \nu_x \,+\, e^-$. Light water detectors 
are mainly sensitive to $\nu_e$, but also to $\nu_{\mu}$ and 
$\nu_{\tau}$, such that
$\sigma(\nu_{\mu\tau} \, e^- \to \nu_{\mu\tau} \, e^-) \simeq 
0.15 \times \sigma(\nu_e \, e^- \to \nu_e \, e^-)$.

The follow-up of the Kamiokande project is called the SuperKamiokande (SK) 
experiment~\cite{bib:SK1}. It was built to investigate in more detail the
nature of atmospheric and solar neutrino oscillations. 
The SK detector is a huge, 40~m in diameter and 40~m
high, circular cylinder filled with 50,000 tons of ultra-pure light water. 
The SK detector operates at an energy threshold that permits the study 
of the $^8B$ neutrinos.
It is divided into an outer detector to veto incoming cosmic
ray muons and to shield external low energy background; and 
an inner detector (32,000 tons, of which 22,500 tons is the active fiducial
volume) viewed by 11,146 PMT. As in Kamiokande, solar neutrinos are observed by
detecting \v{C}erenkov photons emitted by the electrons
resulting from ES events. The event rate was about 15 events per day (substantially
larger than the rate in the radiochemical experiments).
The SK-I data allows measurements of the time dependence of
the ES rate. It led to the measurement of
the day/night rate asymmetry~\cite{bib:SK1}
\begin{equation}
A_{\rm{DN}}
= 2 \frac{\Phi_{\rm{D}}-\Phi_{\rm{N}}}{\Phi_{\rm{D}}+\Phi_{\rm{N}}} 
= -0.021 \pm 0.020 \, ^{+0.013}_{-0.012} \, ,
\end{equation}
and the precise determination of the ES neutrino rate
$\Phi_{\rm{ES}} = (2.35 \pm 0.02 \pm 0.08) \times 10^6~{\rm{cm}}^{-2} {\rm{s^{-1}}}$~\cite{bib:SK1}.
The energy shape of the recoil electron agrees well, within
experimental errors, with that predicted from the neutrino
spectrum from the beta decay of $^8B$. 
The measurement of the absolute flux, however, is about 41\% of
that predicted by the SSM.

The results of the second phase of SK (SK-II) are consistent 
with the results of SK-I. The measured ES neutrino rate is 
$\Phi_{\rm{ES}}= 2.38 \pm 0.05 _{-0.15}^{+0.16} \times 10^6~{\rm{cm}}^{-2} {\rm{s^{-1}}}$~\cite{bib:SK2};
while the day-night difference is found to be $A_{\rm{DN}} = -0.063 \pm 0.042 \pm 0.037$~\cite{bib:SK2}. 

\section{Borexino}

The Borexino detector~\cite{bib:bxdetectorpaper} is located in Hall C of the
Gran Sasso Laboratory. It is
a 300 ton liquid-scintillator based detector with 100 tons of
active fiducial mass in a 8.3~m diameter spherical nylon bag surrounded
by a 2.6 meter thick spherical shell filled with buffer oil.
The liquid scintillator and buffer liquid are viewed
by 2,240 PMT which are mounted inside a
13.5~m diameter stainless steel tank;
which is in turn surrounded by a 18~m spherical tank filled with ultra-pure
light water to act as a radiation shield. 

Solar neutrinos are detected in Borexino
through their elastic scattering on electrons in the scintillator.  Electron 
neutrinos ($\nu_e$) interact through charged and neutral currents and in the energy range of interest have a 
cross section $\sim$5~times larger than $\nu_\mu$ and $\nu_\tau$, which interact only via the neutral current.  
The electrons scattered by neutrinos are detected by means of the scintillation light. 
Borexino is the first solar neutrino experiment to report a real-time observation of low energy $^7Be$
neutrinos. The 
signature for the mono-energetic $^7Be$ neutrinos in Borexino is the Compton-like edge 
of the recoil electrons at 665~keV.

Borexino reported the direct measurement of the 0.862~MeV $^7Be$ solar neutrinos with the Borexino detector 
from an analysis of 192 live days in the period from May 16, 2007 to April 12, 2008, totaling a 41.3~ton$\cdot$yr fiducial 
exposure to solar neutrinos. It yields an 
interaction rate of 49$\pm$3$_{\rm stat}$$\pm$4$_{\rm syst}$~counts/(day$\cdot$100~ton)~\cite{bib:bxBe7}.

\section{Sudbury Neutrino Observatory}

The Sudbury Neutrino Observatory (SNO) was a 1,000 ton heavy-water
\v{C}erenkov detector~\cite{bib:snonim} situated 2~km underground in INCO's 
Creighton mine in Canada. Another 7,000 tons of ultra-pure light water 
was used for support and shielding. The heavy water was in 
an acrylic vessel (12~m diameter and 5~cm thick) viewed by 9,456 PMT
mounted on a geodesic structure 18~m in diameter; all contained within a 
polyurethane-coated
barrel-shaped cavity (22~m diameter by 34~m high).
The SNO detector has been filled with water in May 1999 and data taking ended in 2006.
The solar-neutrino detectors in operation prior to SNO were mainly sensitive 
to the electron neutrino type; while the use of heavy water by SNO allowed the flux of 
all three neutrino types to be measured. Neutrinos from $^8B$ decay in the Sun
were observed in SNO from \v{C}erenkov processes following three reactions:
(i) the charged-current (CC) reaction, specific to electron neutrinos
$d \,+\, \nu_e \to p \,+\, p \,+\, e^-$. This reaction has a Q value of 1.4 MeV and
the electron energy is strongly correlated with the
neutrino energy, providing potential sensitivity
to spectral distortions; (ii) the neutral-current (NC) reaction, equally sensitive to all non-sterile 
neutrino types ($x=e, \mu, \tau$) $\nu_x \,+\, d \to n \,+\, p \,+\, \nu_x$.
This reaction has a threshold of 2.2 MeV and
is observed through the detection of neutrons
by three different techniques in separate phases of
the experiment; and (iii) the elastic-scattering (ES) reaction $\nu_x \,+\, e^- \to \nu_x \,+\, e^-$.
This reaction has a substantially lower cross section than the other two and as mentioned before is 
predominantly sensitive to electron neutrinos.

The relations $\Phi_{\rm{CC}}= \phi_e$, $\Phi_{\rm{ES}}= \phi_e + 0.15 \phi_{\mu\tau}$ and
$\Phi_{\rm{NC}}= \phi_e + \phi_{\mu\tau}$ gave SNO the status of an appearance experiment.
The SNO experimental plan called for three phases
wherein different techniques 
were employed for the detection of neutrons from 
the NC reaction. During the first
phase, with pure heavy water, neutrons were observed 
through the \v{C}erenkov light produced when
neutrons were captured on deuterium, producing
6.25 MeV gammas. For
the second phase, about 2~tons of $NaCl$ was
added to the heavy water and neutron detection
was enhanced through capture on $Cl$, with
about 8.6~MeV gamma energy release.
For the third phase, the salt was removed and an array of
$^3He$-filled proportional counters was installed
to provide direct detection of neutrons.

SNO reported results from a joint analysis of Phase~I and
Phase~II data~\cite{bib:snoleta}.  The effective
electron kinetic energy threshold used is 3.5~MeV, the
lowest analysis threshold yet achieved with water Cherenkov detector
data. Overall, the low threshold
increased the statistics of the CC and ES events by roughly 30\%, and
of NC events by $\sim$70\%.  
In units of $ 10^6$ cm$^{-2}$ s$^{-1}$, a fit in which 
the free parameters directly
describe the total $^8B$ neutrino flux and the energy-dependent
$\nu_e$ survival probability provides a measure of the total $^8B$
neutrino flux $\Phi_{^8B} = 5.046 ^{+0.159}_{-0.152} \, ^{+0.107}_{-0.123}$.
The uncertainty on $\Phi_{^8B}$ have been significantly reduced
compare to previously published results by SNO.
The fit for the survival probability assumes
unitarity of the neutrino mixing matrix, and that the underlying
neutrino spectrum follows a smoothly-distorted $^8B$ shape.
The day survival probability extracted by SNO is parameterized as a second-order polynomial
$P_{ee}^{\rm day}(E_\nu) = c_0 + c_1 (E_\nu - 10\;{\rm MeV}) + c_2 (E_\nu -
10\;{\rm MeV})^2$; while
allowing for a linear energy-dependent asymmetry between day and night
spectra $A(E_\nu) = a_0 + a_1(E_\nu - 10\;{\rm MeV})$. 
The clear deviation from unity of the constant term of the day survival probability 
$c_0 = 0.3435 \, ^{+0.0205}_{-0.0197} \, ^{+0.0111}_{-0.0066} \, ^{+0.0050}_{-0.0059}$ provides
a undeniable signature of solar neutrino oscillations~\cite{bib:snoleta}. On the other hand, 
no evidence for either a significant
spectral distortion or a day/night asymmetry was found.

SNO also used data of Phase~III to measure the rate of NC 
interactions in heavy water and precisely determined the total active ($\nu_{x}$) $^8B$ solar 
neutrino flux~\cite{bib:snoncd}. The total solar neutrino flux is found to be $5.54 ^{+0.33}_{-0.31} \, ^{+0.36}_{-0.34}
\times10^{6}$~cm$^{-2}$s$^{-1}$, in agreement with Phase~I and~II measurements and the SSM.

\section{Global Fits}

Constraints on neutrino mixing parameters can be derived by comparing
neutrino oscillation SSM predictions with experimental data. A
three-flavor, active neutrino oscillation model has four parameters:
$\theta_{12}$ and $\theta_{13}$, which quantify the strength of the
mixing between flavor and mass eigenstates, and $\Delta m^2_{21}$ and
$\Delta m^2_{31}$, the differences between the squares of the masses
of the neutrino propagation eigenstates.  Note that the remaining
mixing angle, $\theta_{23}$, and the CP-violating phase, $\delta$, are
irrelevant for the oscillation analysis of solar neutrino data.
The value of $\Delta m^2_{31}$ was fixed to $+2.3\times 10^{-3}\
\mathrm{eV^2}$.

For each set of parameters, the oscillation model was used to predict
the rates in the Chlorine~\cite{bib:cleveland}, Gallium~\cite{bib:ga} and
Borexino~\cite{bib:bxBe7} experiments, the Super-Kamiokande Phase~I zenith
spectra~\cite{bib:SK1} and Phase~II day/night spectra~\cite{bib:SK2}, 
SNO Phases~I and II survival probability day/night curves~\cite{bib:snoleta}, and the SNO Phase~III 
rates~\cite{bib:snoncd}.  The expected rates and spectra were divided by the
respective predictions, calculated without oscillations, to remove the
effects of the model scaling factors.  The unitless rates were then
used in a global $\chi^2$ calculation. For completeness, the
rates and spectrum measured by the antineutrino reactor experiment KamLAND~\cite{bib:kam}
are added in the global fit to constrain even further the solar neutrino mixing parameters.
The KamLAND rates and spectrum were 
predicted using three-flavor vacuum oscillations.

\begin{figure}[!ht]
\begin{center}
\includegraphics[width=0.49\textwidth]{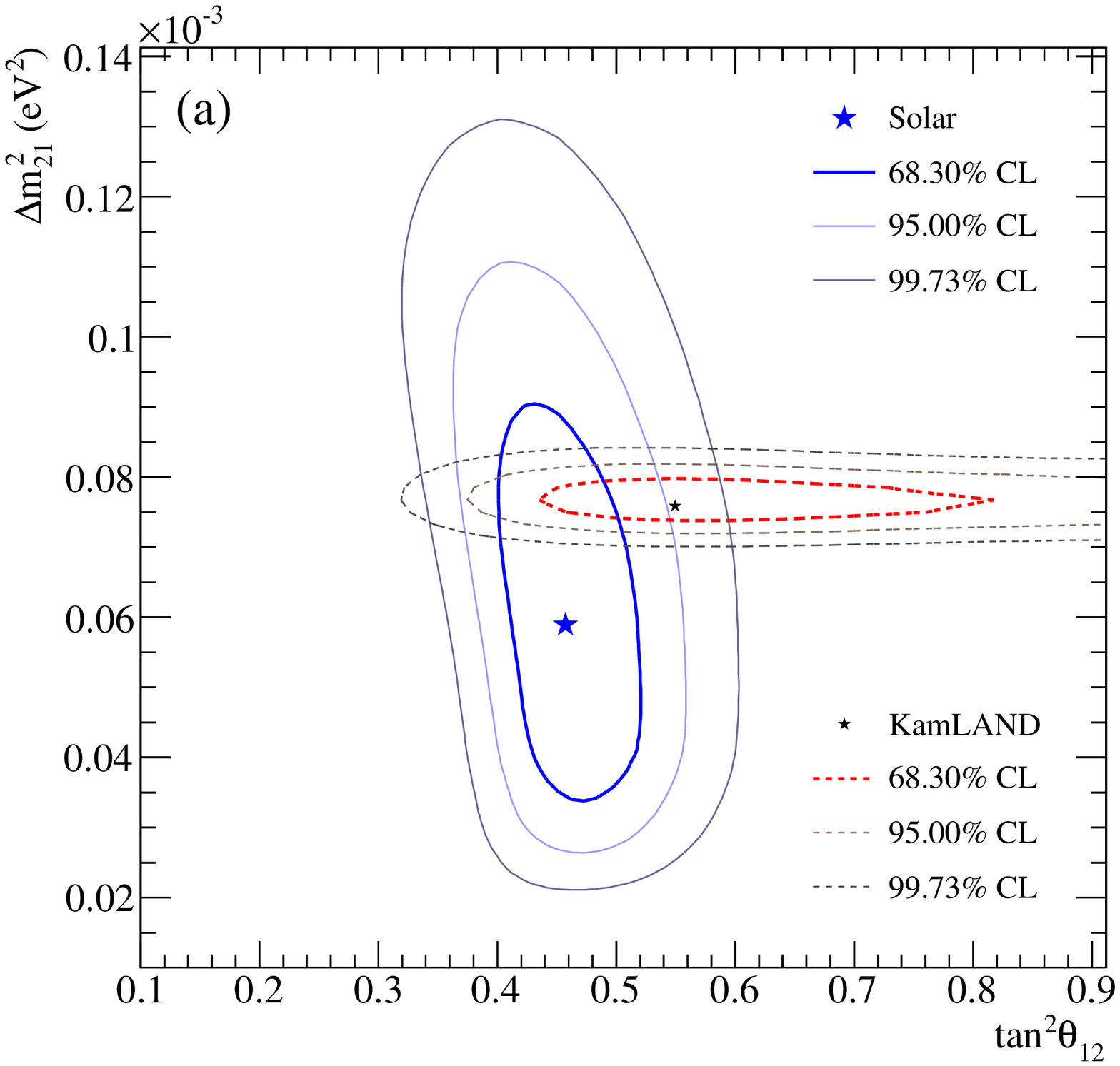}
\includegraphics[width=0.49\textwidth]{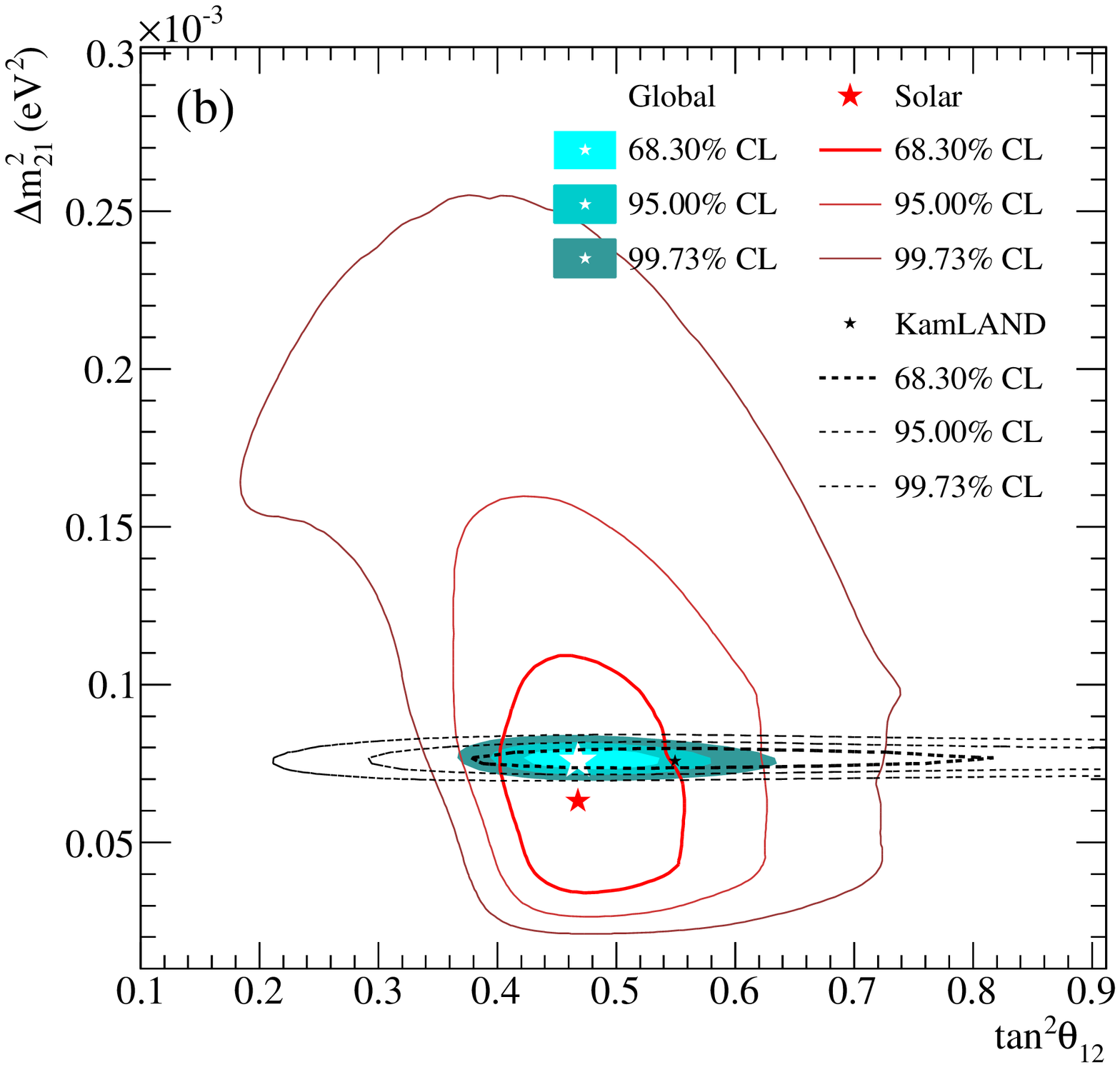}
\caption{Solar and KamLAND oscillation parameter analysis for a) a
two-flavor oscillation hypothesis and b) a three-flavor hypothesis.
The solar data includes Cl, SAGE, Gallex/GNO, Borexino,
SK-I zenith and SK-II day/night spectra, SNO Phase~I+II survival probability day/night
curves and SNO Phase~III integral rates.  The $\chi^2$ is minimized
with respect to all undisplayed parameters, including
$\sin^2\theta_{13}$ and $\Phi_{^8{\rm B}}$~\cite{bib:snoleta}.\label{f:contour-3nu} }
\end{center} 
\end{figure}

Fig.~\ref{f:contour-3nu}(a) shows
an overlay of the global solar and the KamLAND allowed regions in
$\tan^2\theta_{12}$ and $\Delta m^2_{21}$ parameter space, under a
two-flavor hypothesis. Fig.~\ref{f:contour-3nu}(b) shows the same
overlay for the three-flavor hypothesis that allows the value of
$\sin^2\theta_{13}$ to be non-zero. The three-flavor contours show the effect of
marginalizing both $\Phi_{^8{\rm B}}$ and $\sin^2\theta_{13}$ at each
point in space.  

Table~\ref{t:oscpars} summarizes the oscillation parameters from a
two-flavor oscillation analysis, while Table~\ref{t:oscpars3}
summarizes the results from a three-flavor oscillation analysis,
performed in the context of a global fit~\cite{bib:snoleta}.

\begin{table}[!h]
\begin{center}
\begin{tabular}{lcc}
\hline \hline Oscillation analysis & $\tan^2\theta_{12}$ & $\Delta
m^2_{21}(\mathrm{eV}^2)$\\ 
\hline Solar &
$0.457\,^{+0.038}_{-0.041}$ & $5.89\,^{+2.13}_{-2.16}\times
10\,^{-5}$\\ Solar+KamLAND & $0.457\,^{+0.040}_{-0.029}$ &
$7.59\,^{+0.20}_{-0.21}\times 10\,^{-5}$\\ \hline &
$\chi^2_{\mathrm{min}}/\mathrm{ndf}$ & $\Phi_{^8{\rm B}}$ ($\times
10^6\,\rm cm^{-2}\,s^{-1} $)\\ \hline Solar & $67.5/89$ &
$5.104\,^{+0.199}_{-0.148}$\\ Solar+KamLAND & $82.8/106$ &
$5.013\,^{+0.119}_{-0.148}$\\ \hline \hline
\end{tabular}
\caption{Best-fit neutrino oscillation parameters and extracted $^8$B
flux from a two-flavor oscillation analysis. 
Uncertainties listed are $\pm 1\sigma$ after the $\chi^2$ was
minimized with respect to all other parameters.}
\label{t:oscpars}
\end{center}
\end{table}

\begin{table}[!h]
\begin{center}
\begin{tabular}{lcc}
\hline \hline Oscillation analysis & $\tan^2\theta_{12}$ & $\Delta
m^2_{21}(\mathrm{eV}^2)$\\ \hline Solar & $0.468\,^{+0.052}_{-0.050}$
& $6.31\,^{+2.49}_{-2.58}\times 10\,^{-5}$\\ Solar+KamLAND &
$0.468\,^{+0.042}_{-0.033}$ & $7.59\,^{+0.21}_{-0.21}\times
10\,^{-5}$\\ \hline & $\chi^2_{\mathrm{min}}/\mathrm{ndf}$ &
$\Phi_{^8{\rm B}}$ ($\times 10^6\,\rm cm^{-2}\,s^{-1} $)\\ \hline
Solar & $67.4/89$ & $5.115\,^{+0.159}_{-0.193}$\\ Solar+KamLAND &
$81.4/106$ & $5.087\,^{+0.171}_{-0.159}$\\ \hline
&\multicolumn{2}{c}{$\sin^2\theta_{13}(\times 10\,^{-2})$} \\ \hline
Solar & \multicolumn{2}{c}{$< 8.10\, (95\%\, {\rm\, C.L.)}$}\\
Solar+KamLAND& \multicolumn{2}{c}{$2.00\,^{+2.09}_{-1.63}$}\\ \hline
\hline
\end{tabular}
\caption{Best-fit neutrino oscillation parameters and extracted $^8$B
flux from a three-flavor oscillation analysis.  Uncertainties listed
are $\pm 1\sigma$ after the $\chi^2$ was minimized with respect to all
other parameters.}
\label{t:oscpars3}
\end{center}
\end{table}

\section{News and Near Future}

The new Borexino measurement~\cite{bib:bxB8} of $\nu$-$e$ elastic scattering  from $^8B$ solar neutrinos 
with a 3\,MeV energy threshold is 
not included in the global fit of the previous section. This result is limited by statistics, but it 
is nevertheless relevant because it is probing low energy $^8B$ neutrinos.
The rate of solar neutrino-induced electron scattering events above this energy 
is $0.217\pm 0.038 \pm 0.008$~counts/(day$\cdot$100~ton),
which corresponds to $\Phi^{\rm ES}_{\rm ^8B}$ = {2.4 $\pm$ 0.4$\pm$ 0.1}$\times$10$^6$~cm$^{-2}$ s$^{-1}$, in good 
agreement with measurements from SNO and SuperKamiokande.  

Preliminary solar neutrino results of the third phase of the SuperKamiokande
presented at this conference~\cite{bib:skposter} have already been submitted
for publication~\cite{bib:SK3}.
With improved detector calibrations, 
a full detector simulation, and improved analysis methods, 
the systematic uncertainty on the total neutrino flux
has been reduced to $\pm 2.1\%$. This leads to a significant contraction of 
the systematic uncertainty compared to the first phase.
The energy threshold of $^8B$ neutrino events of SK-III is also pushed lower
with respect to SK-I. It yields an ES rate of
2.32 $\pm$ 0.04 $\pm$ 0.05 $\times 10^{6}$~cm$^{-2}$sec$^{-1}$,
in agreement with previous measurements.

It is clear that the uncertainty on the solar mixing angle $\theta_{12}$ has been
noticeably reduced with data analyses that use lower energy thresholds since 
it enhanced the experimental ability to explore the
SSM prediction of energy-dependent matter-induced neutrino oscillation. Very soon, Borexino and SNO
will release new results on $^7Be$ and $^8B$ neutrinos, respectively.
Borexino reported at ICHEP2010 a precise day/night asymmetry measurement 
$A_{\rm{DN}} = 0.007 \pm 0.073$~\cite{bib:bxADN}.
It is foreseen that the statistical precision of Borexino will allow
for a detailed investigation of mixing solutions.
Expect further improvement with SNO's final joint model-independent fit of the solar survival probability
with data from Phases~I, II and III.

%

\section{Summary}

Solar neutrino oscillation is clearly established by the combination of the 
results from the chlorine, gallium, SK, Borexino and SNO
experiments. The real-time data of SK, Borexino and SNO do not show large energy distortion 
nor time-like asymmetry.
SNO provided the first direct evidence of flavor conversion of solar electron 
neutrinos by comparing the
CC and NC rates. Matter effects seem to explain the energy dependence of solar 
oscillation, and Large Mixing Angle (LMA) solutions are favored.

After 30 years of hard labor from the nuclear and particle physics 
community, the Solar Neutrino Problem is now an industry for 
precise measurements of neutrino oscillation parameters with the next 
generation of solar neutrino and long baseline neutrino experiments
at the horizon.

\section*{Acknowledgments}
This article builds upon the careful and detailed work of
many people. 
The author has been financially supported in
Canada by the Natural Sciences and Engineering Research 
Council (NSERC), the Canada Research Chair (CRC) Program,
and the Canadian Foundation for Innovation (CFI).

\end{document}